\documentclass[
    ,final            
  ]
  {aipproc}

\layoutstyle{6x9}

\begin{document}

\title{First Results from IceCube}



\author{Spencer R. Klein, for the IceCube 
Collaboration\footnote{The complete author list is available at 
http://icecube.wisc.edu/pub\_and\_doc/conferences/panic05.}}
{address={LBNL, 1 Cyclotron Rd., Berkeley, CA, 94720, USA}}

\begin{abstract}

IceCube is a 1 km$^3$ neutrino observatory being built to study
neutrino production in active galactic nuclei, gamma-ray bursts,
supernova remnants, and a host of other astrophysical sources.
High-energy neutrinos may signal the sources of ultra-high energy
cosmic rays.  IceCube will also study many particle-physics topics:
searches for WIMP annihilation in the Earth or the Sun, and for
signatures of supersymmetry in neutrino interactions, studies of
neutrino properties, including searches for extra dimensions, and
searches for exotica such as magnetic monopoles or Q-balls.  IceCube
will also study the cosmic-ray composition.

In January, 2005, 60 digital optical modules (DOMs) were deployed in
the South Polar ice at depths ranging from 1450 to 2450 meters, and 8
ice-tanks, each containing 2 DOMs were deployed as part of a surface
air-shower array.  All 76 DOMs are collecting high-quality data. After
discussing the IceCube physics program and hardware, I will present
some initial results with the first DOMs.

\end{abstract}

\maketitle

\section{IceCube Physics Program}

Neutrinos are strongly penetrating particles that can be used to
search for the sources of high-energy cosmic rays and study many
energetic astrophysical phenomena.  Except at the very highest
energies (above 50 EeV), high-energy protons and nuclei are bent in
the interstellar magnetic fields, and high-energy photons are absorbed
by interactions with the $3^0$K microwave background and infrared
photons from red-shifted starlight.  Only neutrinos can provide a
clear view of the sky above 50 TeV.  The primary purpose of IceCube is
to map out this sky \cite{PDD}; many calculations using a wide variety
of approaches find that a volume of at least 1 km$^3$ detector is
required to observe neutrino sources \cite{icsense}.  Neutrinos are
expected as `by-products' of hadronic acceleration, and may be
observed from active galactic nuclei (AGNs), gamma-ray bursters
(GRBs), and possibly supernova remnants.

The completed IceCube detector will collect data on about 100,000
atmospheric $\nu$/year \cite{atmosphere}.  These neutrinos can be used
to measure the neutrino-nucleon cross sections (through absorption in
the earth) and search for new forms of neutrino oscillations,
such as those predicted by some models of quantum gravity \cite{qg}).

IceCube will also search for signs of supersymmetry (some parameter
sets with high mass scales lead to pairs of upward going particles in
the detector) \cite{SUSY}, neutrinos from WIMP annihilation in the
Earth or the Sun \cite{WIMPs}, and for exotic particles like magnetic
monopoles, Q-balls, and the like.  IceCube also serves as a supernova
monitor; a supernova in our galaxy should increase the singles rates
of all of the in-ice phototubes \cite{supernova}.

\section{Hardware overview}

IceCube is building on the experience of AMANDA, which studied a
variety of neutrino topics \cite{paolo}.  IceCube will instrument 1
km$^3$ using 4800 digital optical modules (DOMs).  The DOMs will be
deployed in 80 strings on a 125 meter hexagonal grid. Each string will
contain 60 modules at a 17 meter spacing.

Each DOM acts like a mini-satellite, acquiring data autonomously.  A
DOM consists of a 13'' pressure vessel (Benthosphere) containing a
Hamamatsu R7081-02 10'' photomultiplier tube (PMT), PMT base, flasher
(LED) board and a main electronics board.  The main board, Fig. 1,
includes front-end electronics and trigger, two analog-to-digital
converter systems, a precision clock and a large field programmable
gate array (FPGA), with an integrated CPU, for control,
communications, etc \cite{stokstad}.  All communication with the
surface is through a single twisted-wire pair (shared by 2 DOMs); this
single pair carries power, bi-directional data, and timing calibration
signals.

The PMT gains are currently set at $10^{7}$.  The PMT signal is AC
coupled through a transformer on the base.  The signal is fed to a
discriminator and two separate digitizer circuits.  The discriminator
is typically set at about 1/3 of a photoelectron pulse.  When the
discriminator fires, a digitization cycle begins on the next clock
transition.

\begin{figure}
  \includegraphics[height=.35\textheight]{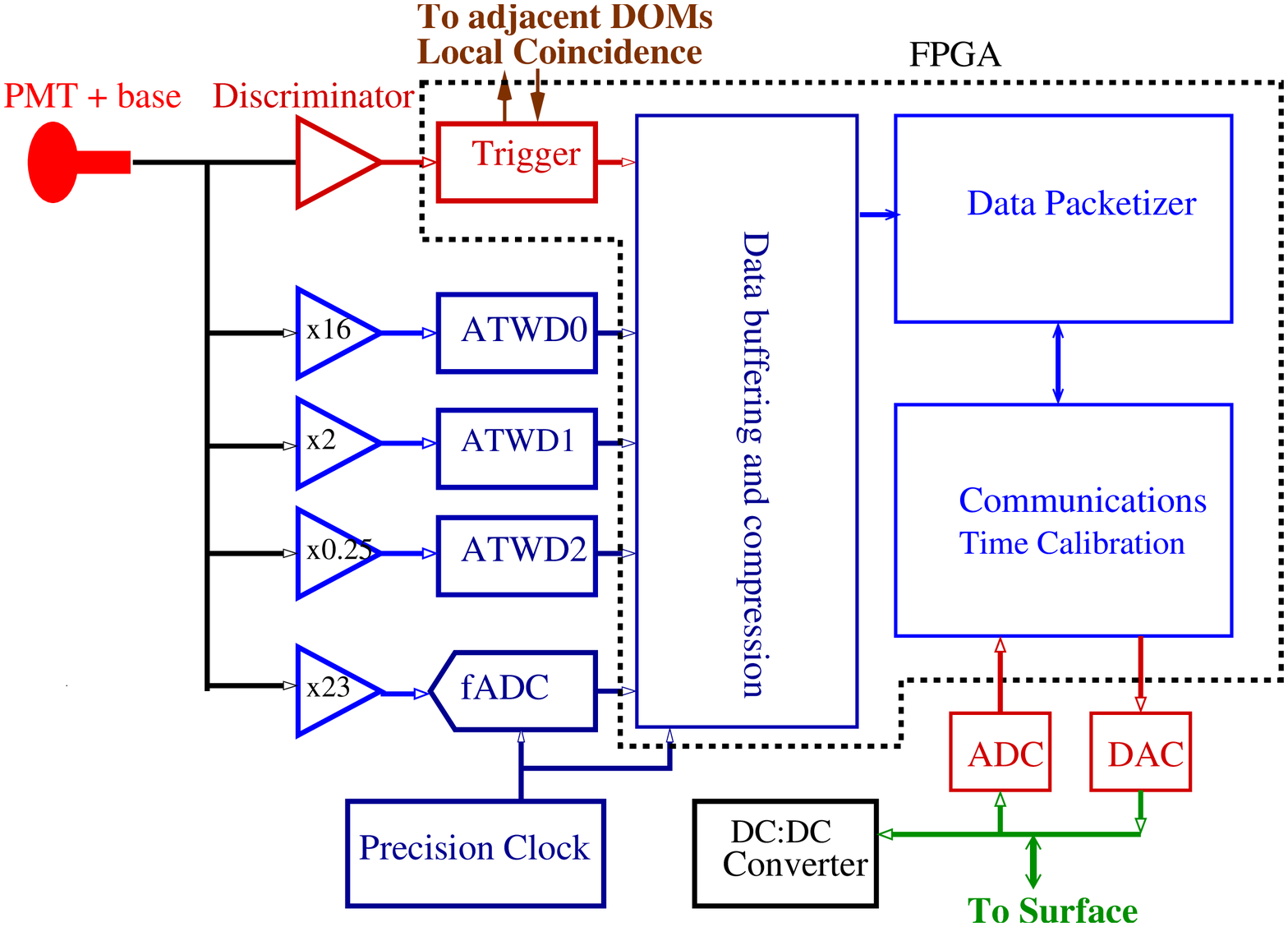}
  \caption{A block diagram of an IceCube DOM main board.}
\end{figure}

The first digitizer is based on a custom switched-capacitor-array
chip, the Analog Transient Waveform Digitizer (ATWD).  It takes 128
samples at from 200 to 700 megasamples per second (MSPS), and then
digitizes them with 10-bit resolution.  IceCube is currently running
at 300 MSPS.  The PMT signal feeds 3 ATWD channels, with a nominal
gain ratio of 0.25:2:16; combined, the three channels offer 14 bits of
resolution, covering from single photoelectrons up to the maximum PMT
voltage.  To reduce dead time, each DOM contains two ATWDs which
operate in a ping-pong fashion. A precision 20 MHz clock is a local
reference, used to relate the ATWD launch time to global time.

The other digitizer system detects late arriving light which has
scattered in the ice.  It uses a 10-bit 40-MSPS commercial ADC to
record signals arriving up to 6.4$\mu$s after the DOM launch.  The ADC
input signals are shaped to ensure that single photoelectrons are
observable.

A local coincidence circuit can select which data is transmitted to
the surface.  Adjacent DOMs are connected by cables, and DOMs can be
configured to react differently to events where adjacent (or
next-to-nearest neighbor) DOMs are also hit.

The DOMs FPGA controls the data acquisition and storage, compresses
the data, packetizes it, and transmits the packets to the surface.  It
controls and calibrates the system, monitors temperatures and
voltages, and monitors the discriminator singles rate; the latter also
serves as a monitor to search for supernovae.  The DOM communicates
with the surface via analog-to-digital and digital-to-analog
converters connected to the twisted pairs.

The DOM clock is a precision 20 MHz oscillator, with a short-term
frequency stability (Allen variance) of $\delta\!f\!/\!f < 10^{-10}$. To
maintain accurate inter-DOM timing, the clocks are periodically
(currently once every 2.5-3.5 seconds) calibrated by a process known
as reciprocal active calibration (RapCal).  A pulse is sent from the
surface down to the DOM.  The signal is received, held for a specified
time, and then retransmitted to the surface.  By using identical DACs
and ADCs at both the DOM and the surface, the transmission time will
be the same in both directions; by comparing the surface and DOM
clocks, an accurate calibration can be maintained.  Data (discussed
below) show that the system maintains timing across the entire array
to better than 2 nsec, considerably below the design requirement of 5
nsec \cite{chirkin}.

The DOM ``flasher boards'' hold 12 light emitting diodes (LEDs) spaced
around the DOM.  Half of the LEDs point horizontally outward, while the
other half point upwards at $45^o$.  These LEDs are used for
calibration, and can mimic $\nu_e$ induced showers.  The LEDs are
individually switchable, and the overall pulse width and amplitude
and programmable.

On the surface, the DOM cables are connected to Digital Receiver
cards, which plug into PCI slots in off-the-shelf computers.  $\pm$ 48
volt power supplies power the DOMs through the DOR card.  A software
trigger uses data from all of the DOMs to select interesting time
regions (events) for storage to disk and/or satellite transmission to
the North. The basic algorithm uses the hit multiplicity in a
specified time window; future algorithms may also examine the topology
and launch times of the hit DOMs.  A GPS clock provides absolute
timing information.

The IceTop surface air-shower array will consist of 160 tanks, two
near the top of each in-ice string.  Each consists of 2 DOMs frozen
into a 2.7 m diameter ice-filled tank.  The tanks are sensitive to
both muons and electromagnetic particles (electrons and photons) in
cosmic-ray air showers.  IceTop will be used as an air-shower array,
as a calibration aid for the in-ice array, and as a veto for IceCube.
By using the in-ice detectors as a veto, it may also be used to search
for high energy gamma rays.  As an air-shower array, the system will
have a threshold of about 320 TeV.

\section{The 2004/2005 deployment}

IceCube deployment began with the assembly of the hot water drill in
Dec. 2004.  A 5-megawatt heater pumps 750 liters/minute of water to
the drill head; during 2004/5, the drill reached speeds of about 1
meter/minute; the complete hole drilling (including going through the
`firn' region of packed snow), took about 50 hours; in 2005/6 this
time is being reduced significantly.

The first IceCube string was deployed in about 16 hours on Jan. 18,
2005.  After deployment, it took a few days for the water around most
of the DOMs to freeze.  For the deepest DOMs, temperatures are
higher, so freeze-in required a few weeks.

Logistics are a critical issue for IceCube.  The `summer' season (the
only time transportation is available, and when active work is
possible) runs from late October until February 14th.  Even then,
construction is heavily constrained by airlift capacity; everything
from people to fuel must fit on ski-equipped LC-130 turboprops.

\section{Results from the first string}

\begin{figure}
  \includegraphics[height=.25\textheight]{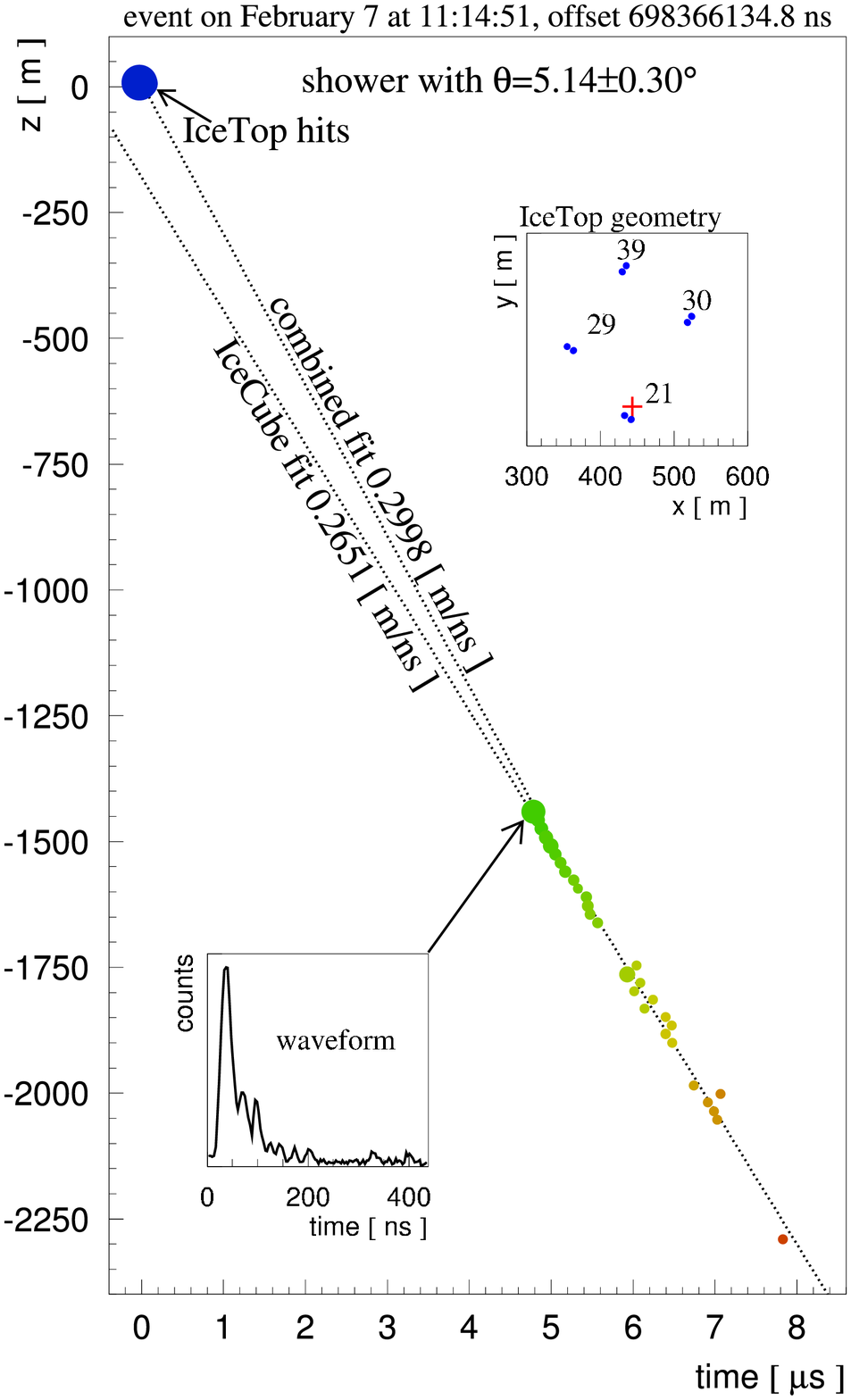} 
  \includegraphics[width=.47\textwidth]{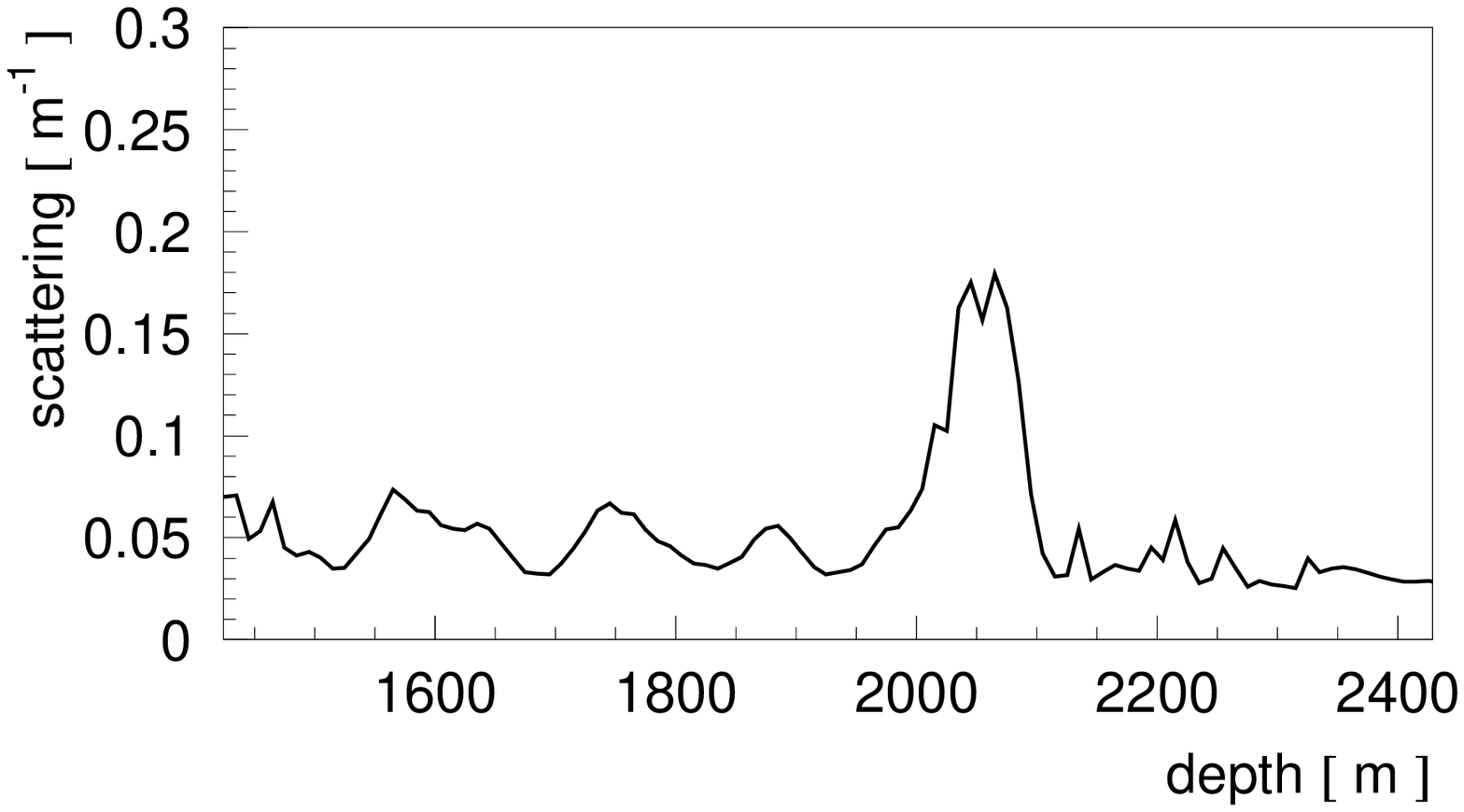} 
\caption{(left) An event that triggered both the IceTop and in-ice
subdetectors.  For the in-ice detectors, the colors indicate the time
that the DOM fired, from green to yellow.  One inset shows on an ATWD
waveform from one of the DOMs; the other shows the IceTop geometry.
The two fits are with and without the IceTop data. (right) The inverse
scattering distance for light (averaged over the IceCube spectral
range) as a function of depth.}
\label{fig:event}
\end{figure}

Most of the results to-date have been obtained studying cosmic-ray
muons and using the flasher boards.  Figure \ref{fig:event} shows an
event that triggered both the in-ice and IceTop detectors.  The four
circles at the top of the picture show the IceTop stations.  The air
shower was reconstructed using a plane wave fit to the IceTop
stations; the in-ice muon was reconstructed with a maximum likelihood
fit that accounted for the depth-dependent scattering of the Cherenkov
photons in the ice.  Figure \ref{fig:event} (right) shows the inverse
scattering distances used. Because only a single string is present,
the azimuthal angle of the muon cannot be determined.

\begin{figure}
  \includegraphics[width=.47\textwidth]{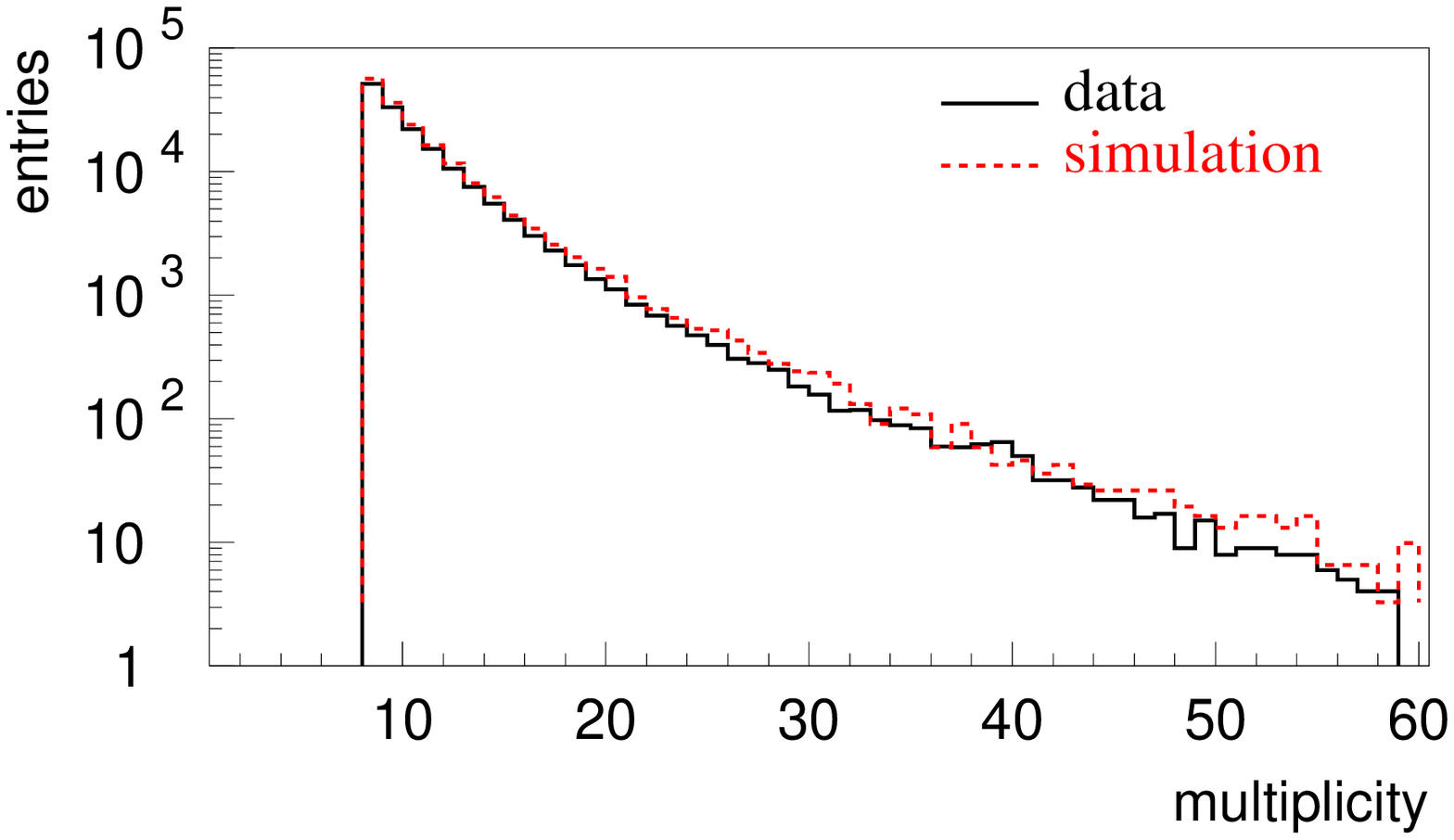}
  \includegraphics[width=.47\textwidth]{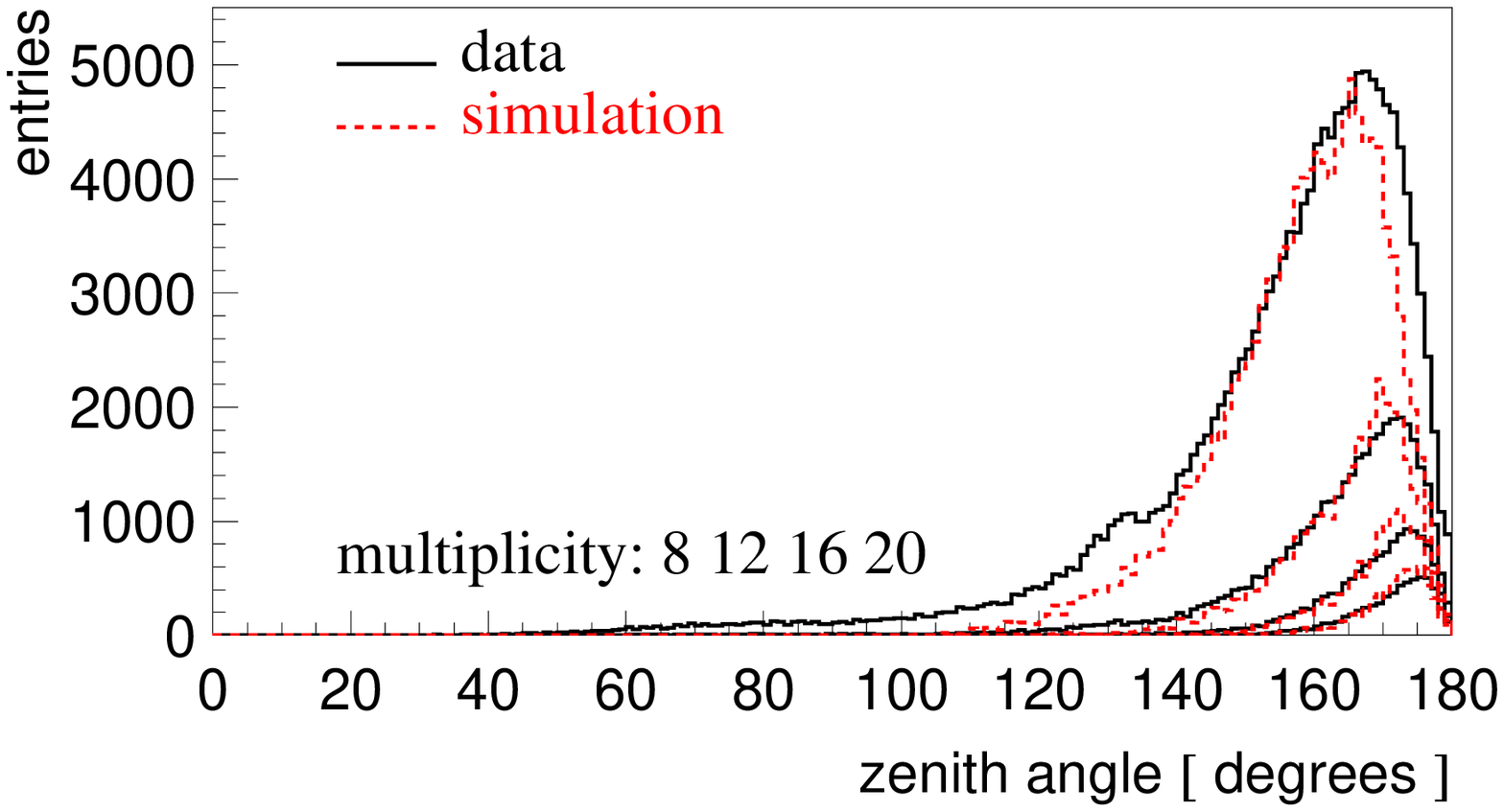} 

\caption{(left) The number of hit DOMs in muon tracks, compared with
simulation.  At least 8 hits are required for a trigger. (right) The
zenith angle distribution for muons events with multiplicity at least
8, 12, 16 or 20 hits, compared with the corresponding simulations.
Here 180$^0$ points straight downward.}
\label{fig:muons}
\end{figure}

The left panel of Figure \ref{fig:muons} shows the number of hit DOMs
in reconstructed muon tracks, while the right panel shows the zenith
angle distribution for reconstructed muons of different minimum
multiplicities, compared with simulations.  The Monte Carlo used here
did not include simulate the ATWD waveforms, so some deviation is to
be expected.

\begin{figure}
  \includegraphics[width=.47\textwidth]{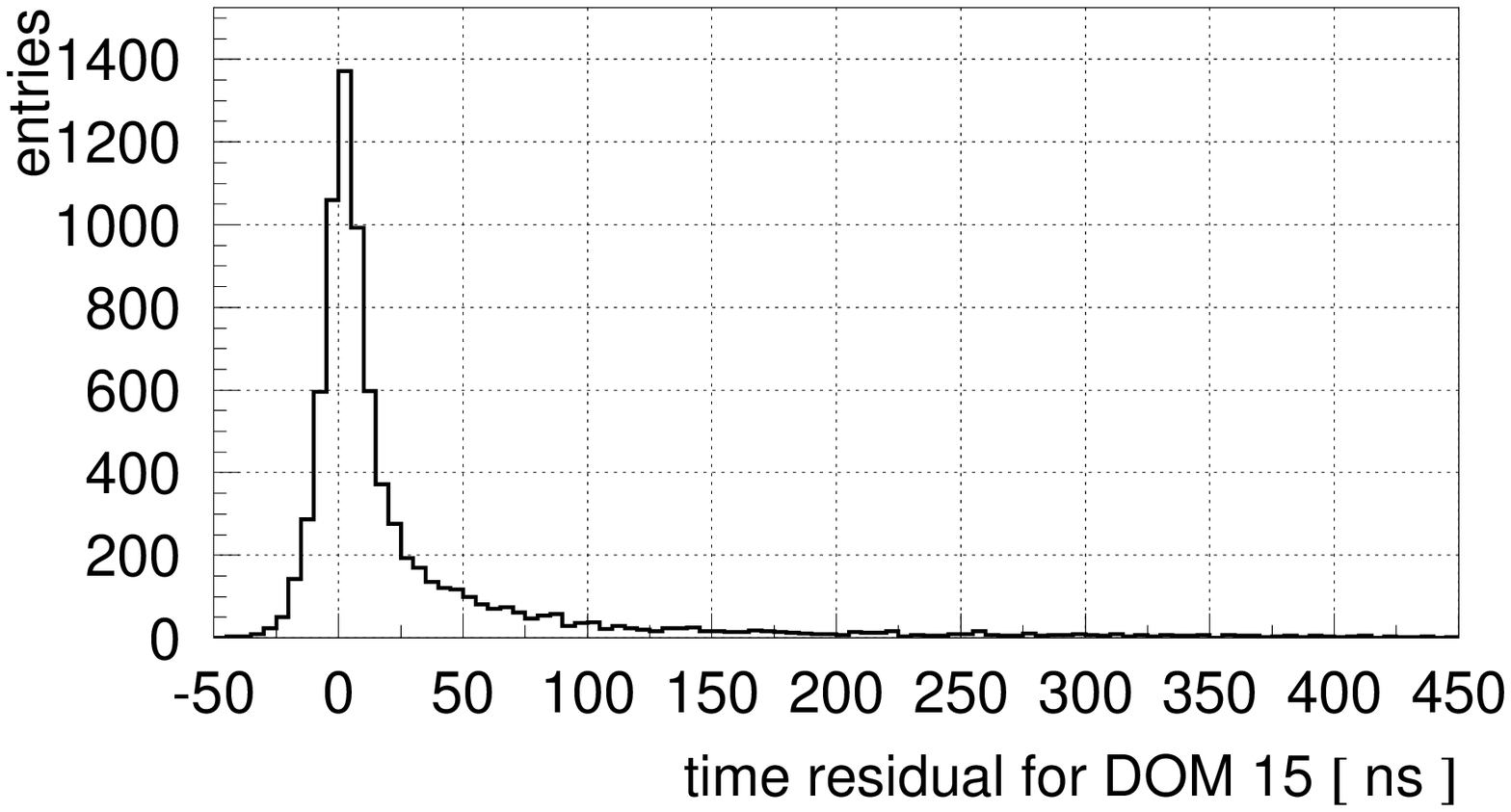}
  \includegraphics[width=.47\textwidth]{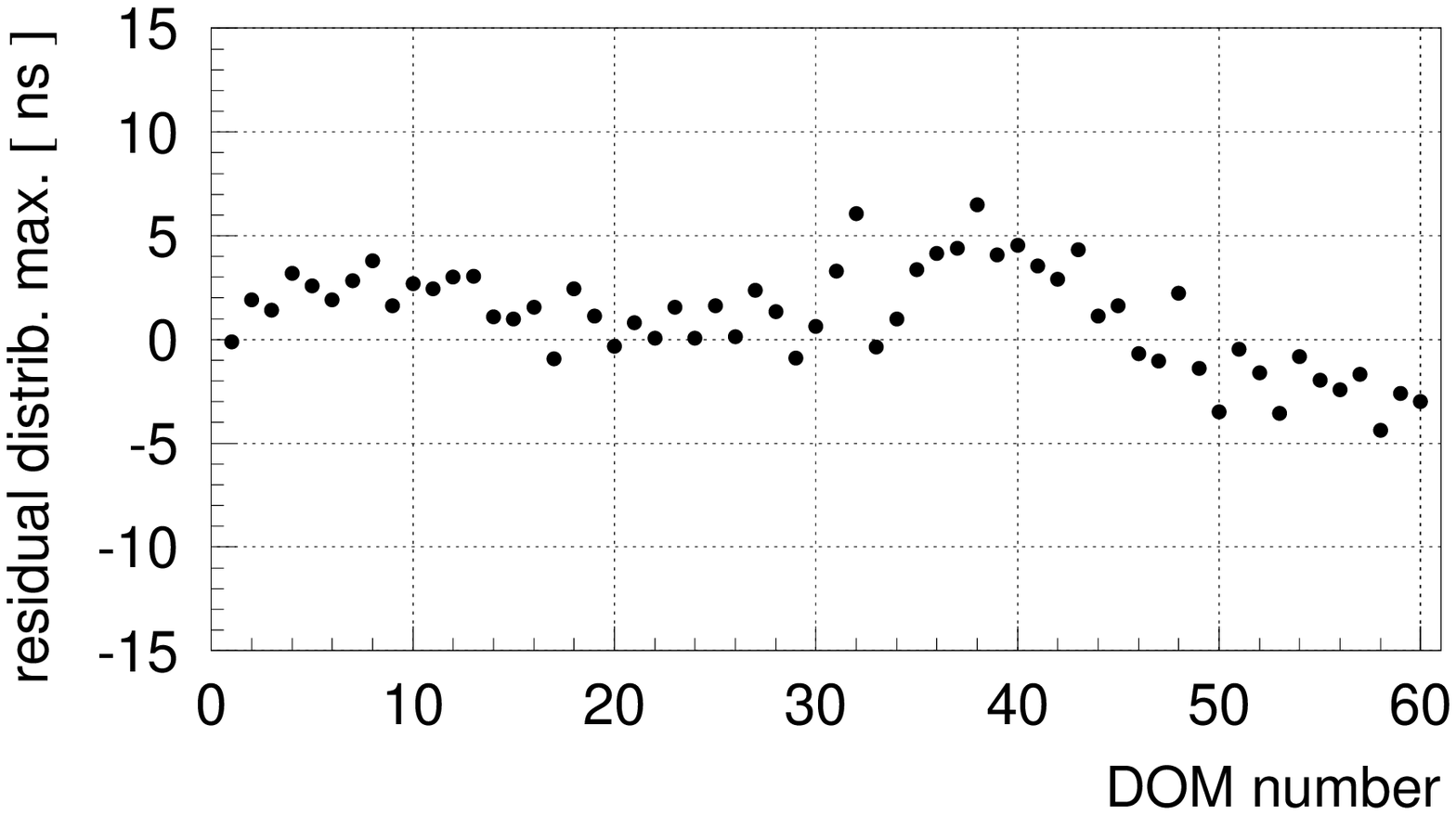} 

\caption{(left) The distribution of timing residuals for a muon track.
Time 0 corresponds to that expected for direct light from the track;
most of the width of the peak comes from errors in the fit,
particularly the distance from the muon to the DOM.  (right) The mean
of the distributions of residuals for all 60 DOMs.}
\label{fig:muontiming}
\end{figure}

Muon tracks are used to verify the detector time resolution.  The muon
track is fit with a given DOM excluded, and then the residual to that
DOM is histogrammed. Figure \ref{fig:muontiming} (left) shows these
residuals for a DOM; the long asymmetric tail is due to light
scattering in the ice, which delays the photons.  The effect of
scattering is minimized by using only muon tracks which pass close to
the DOM whose timing is being measured.  Fig. \ref{fig:muontiming}
(right) shows the mean residuals when this procedure is applied to all
60 DOMs.  As Fig. \ref{fig:event} (right) shows, DOMs 35-45 are in
particularly dusty ice, with a short scattering length; this
scattering causes the means to rise above 0.  Still, these muon fits
show that the DOM timing is correct to within 3 nsec.

\begin{figure}
  \includegraphics[height=.13\textheight]{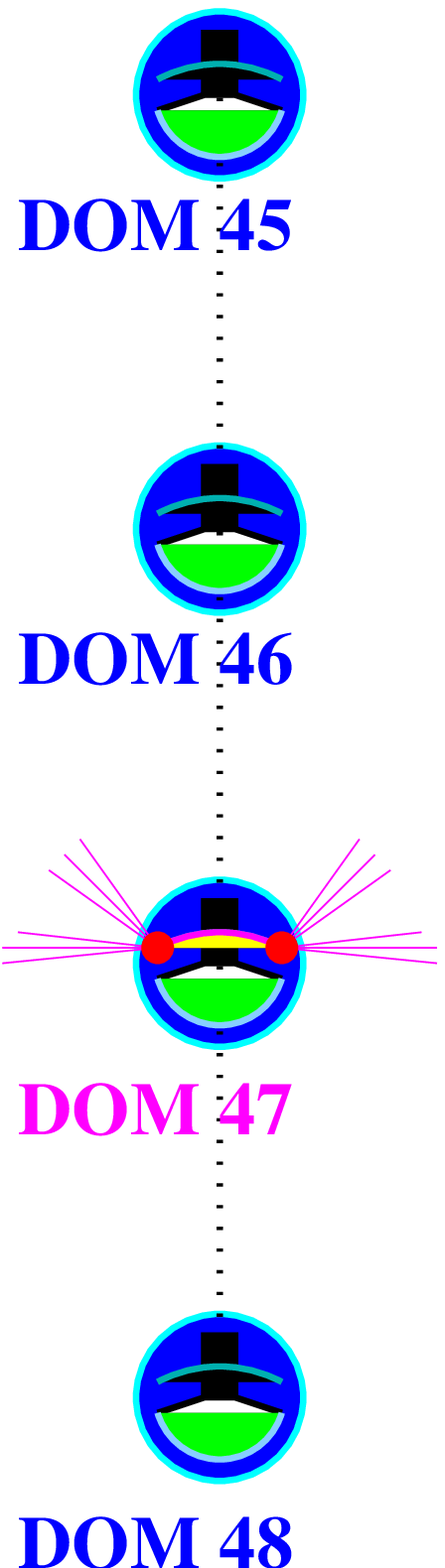}
  \includegraphics[width=.43\textwidth]{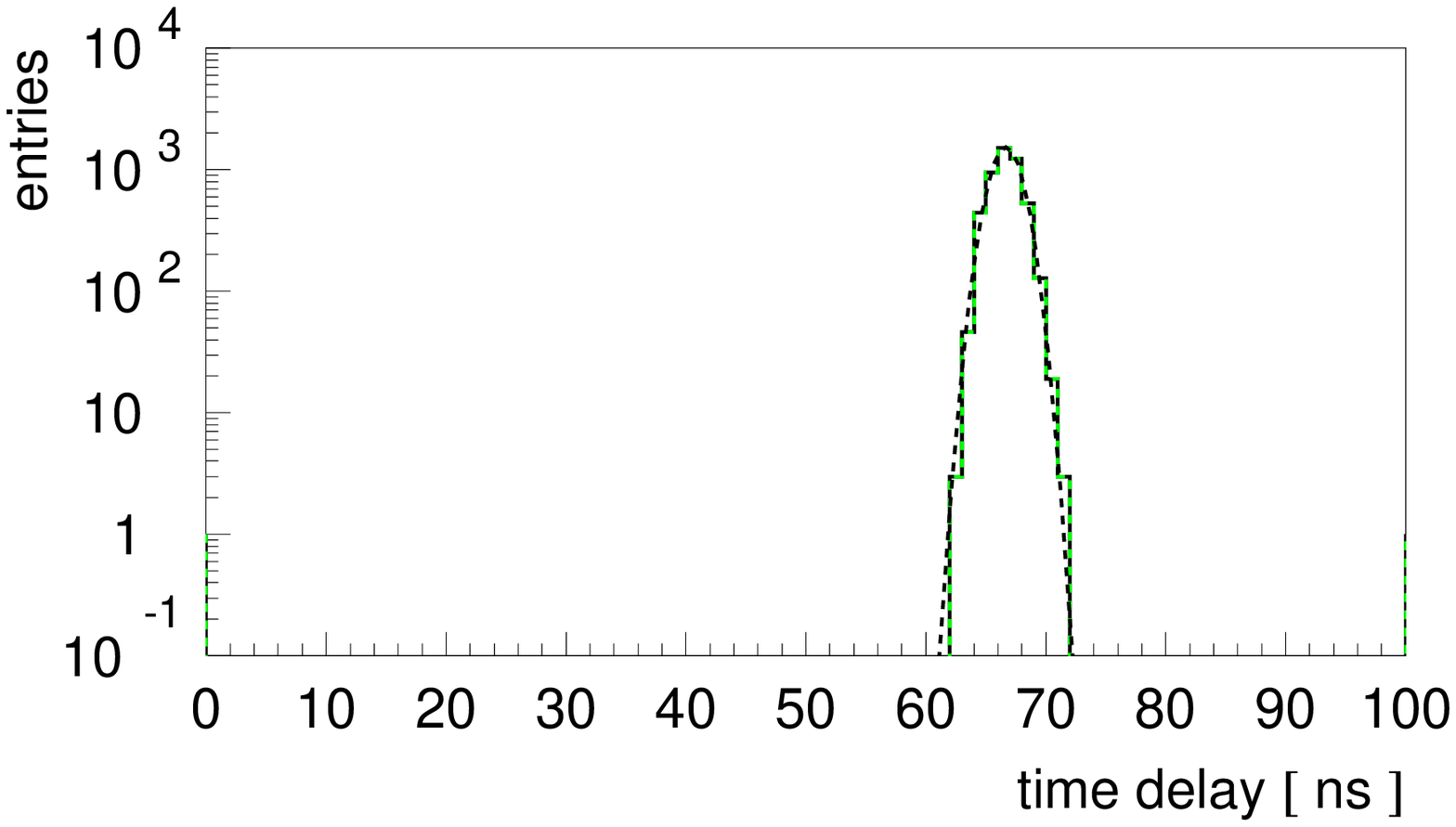}
  \includegraphics[width=.41\textwidth]{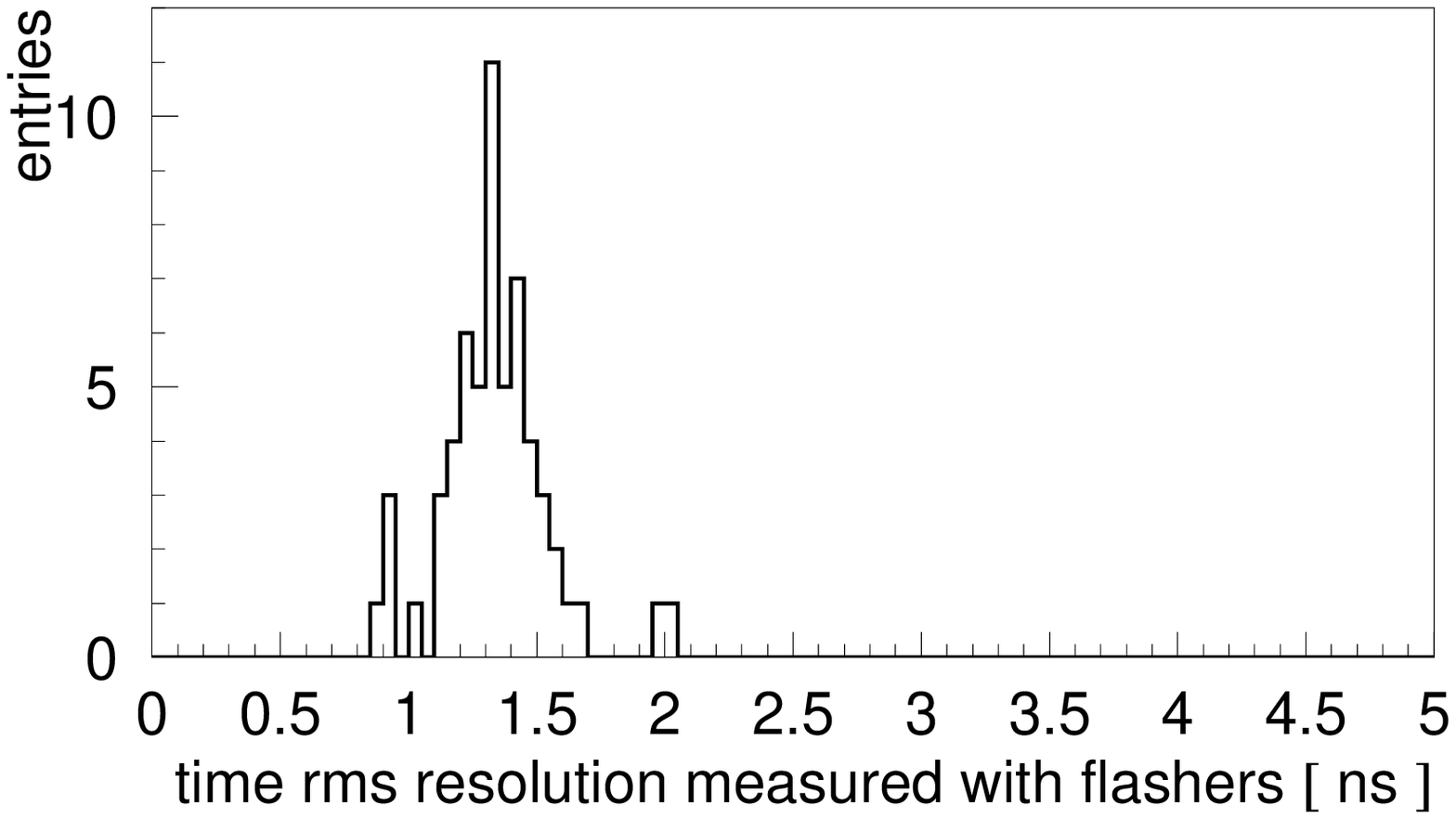} 

\caption{(left) DOM 47 is flashing. (center) The first photon
arrival time difference between DOMs 45 and 46. (right) 
The arrival time differences for the 58 pairs of adjacent DOMs
above a flashing DOM.}
\label{fig:flasher}
\end{figure}

The flasher-board LEDs are also used to study detector performance.
 Figure \ref{fig:flasher} (center) shows the measured time difference
 between DOMs 45 and 46 when DOM 47 is flashing (as shown at the far
 left of Fig. \ref{fig:flasher}); the sigma of the distribution is
 1.26 nsec.  The left histogram shows the measured time differences
 for the 58 adjacent DOM pairs; all the pairs have sigmas under 2
 nsec.

\section{Conclusions}

In 2004/5 the IceCube collaboration deployed a string of 60 DOMs at
the South Pole, along with 8 ice-tanks containing an additional 16
DOMs.  All 76 DOMs are working well, and the system shows time
resolutions of $\approx 2$ nsec.  In 2005/6, we expect to deploy an
additional 8-12 strings, moving toward completion of the 80-string
array by 2010.

\begin{theacknowledgments}

This research was supported by the Deutsche Forschungsgemeinschaft
(DFG); German Ministry for Education and Research; Knut and Alice
Wallenberg Foundation, Sweden; Swedish Research Council; Swedish
Natural Science Research Council; Fund for Scientific Research
(FNRS-FWO), Flanders Institute (IWT), Belgian Federal Office for
Scientific, Technical and Cultural affairs (OSTC), Belgium. UC-Irvine
AENEAS Supercomputer Facility; University of Wisconsin Alumni Research
Foundation; U.S. National Science Foundation, Office of Polar
Programs; U.S. National Science Foundation, Physics Division;
U.S. Department of Energy.

\end{theacknowledgments}


\begin{thebibliography}{9}

\bibitem{PDD}The IceCube Collaboration, {\it Preliminary
Design Document}, 2001.  Available from 
http://icecube.wisc.edu/pub\_and\_doc/.

\bibitem{icsense}J. Ahrens {\it et al.}, Astropart. Phys.
{\bf 20}, 507 (2004).

\bibitem{atmosphere}M. C. Gonzalez-Garcia, F. Halzen and M.
Maltoni, Phys. Rev. {\bf D71}, 093010 (2005).

\bibitem{qg} L. Anchiordoqui {\it et al.}, Phys. Rev. {\bf D72},
065019 (2005).

\bibitem{SUSY}I. F. Albuquerque, G. Burdman and Z. Chako,
Phys. Rev. Lett. {\bf 92}, 221802 (2004).

\bibitem{WIMPs}I. F. Albuquerque, J. Lamoreaux and G. Smoot,
Phys. Rev. {\bf D66}, 125006 (2002).

\bibitem{supernova}A. S. Dighe {\it et al.}, JCAP {\bf 0306}, 005 (2003).

\bibitem{paolo}P. Desiati, these proceedings.

\bibitem{stokstad}R.G. Stokstad, for the IceCube Collaboration,
presented at {\it 11th Wkshp on Electronics for LHC and Future
Experiments}, Sept. 12-16, 2005, Heidelberg, Germany.

\bibitem{chirkin}D. Chirkin, in astro-ph/0509330.

\end{thebibliography}
\end{document}